\documentclass[epj,referee,onecolumn]{svjour}
\usepackage{graphics}
\usepackage{amssymb}
\usepackage{color}

\begin{document}

\title{Quantum quenches in one-dimensional gapless systems} 

\author{Emanuele Coira\inst{1,2} \and Federico Becca,\inst{3} 
\and Alberto Parola\inst{1}}

\institute{
Dipartimento di Scienza e Alta Tecnologia, Universit\`a dell'Insubria, Via Valleggio 11, I-22100 Como, Italy \and
Department de Physique Theorique, University of Geneva, CH-1211 Geneva, Switzerland \and
Democritos Simulation Center CNR-IOM Istituto Officina dei Materiali, Via Bonomea 265, I-34136, Trieste, Italy}

\date{Received: date / Revised version: date}

\abstract{
We present a comparison between the bosonization results for quantum quenches
and exact diagonalizations in microscopic models of interacting spinless 
fermions in a one-dimensional lattice. The numerical analysis of the long-time
averages shows that density-density correlations at small momenta tend to a 
non-zero limit, mimicking a thermal behavior. These results are at variance
with the bosonization approach, which predicts the presence of long-wavelength 
critical properties in the long-time evolution. By contrast, the numerical
results for finite momenta suggest that the singularities at $2k_F$ in the
density-density correlations and at $k_F$ in the momentum distribution are
preserved during the time evolution. The presence of an interaction term that 
breaks integrability flattens out all singularities, suggesting that the time 
evolution of one-dimensional lattice models after a quantum quench may differ 
from that of the Luttinger model.
}

\authorrunning{E.~Coira, F.~Becca, and A.~Parola}
\titlerunning{Quantum quenches in one-dimensional gapless systems} 
\maketitle

\section{Introduction}\label{sec:intro}

The study of non-equilibrium dynamics of {\it isolated} many-body quantum 
systems has been triggered by the recent progress in ultra-cold gases
experiments~\cite{greiner2002,kinoshita2006,hofferberth2007,trotzky2011}.
In fact, these systems are sufficiently weakly coupled to the external 
environment and, therefore, the observation of essentially unitary 
non-equilibrium time evolution on long time scales is possible.
The availability of experimental controllable systems, whose properties 
can be accurately described by simple models, provides an unprecedented 
opportunity to explore new frontiers in physics, including non-equilibrium 
dynamics in closed interacting quantum systems~\cite{polkovnikov2011}.
The experimental advances posed serious challenges to the theory and new 
paradigms must be developed. Although we achieved a satisfactory understanding
of correlated materials at equilibrium, the basic principles governing 
quantum systems far from equilibrium are still in their infancy.

A common protocol to study out-of-equilibrium problems is called
{\it quantum quench} and consists in preparing the system in the ground 
state of a given Hamiltonian and then suddenly let it evolve under the action
of a new Hamiltonian. Since the evolution is unitary, the energy stored into 
the initial state is conserved during the dynamics. The interest in these 
classes of non-equilibrium problems relies on both the dynamics 
itself~\cite{barmettler2009} and the long-time properties, including 
the highly debated issue of thermalization~\cite{rigol2008}.
Quantum quenches have been the subject of vast literature focusing on different
systems~\cite{polkovnikov2011}. In particular, one-dimensional (1D) models
have been largely explored~\cite{cazalilla2011}. From the theoretical point 
of view, there are many different analytical or numerical methods that may 
give useful insights into the dynamical properties and the nature of the 
steady state (if any), inquiring the possibility to reach 
thermalization~\cite{cazalilla2006,kollath2007,manmana2007,calabrese2011}.
Integrable models require a separate analysis~\cite{integrability}, because 
a complete thermalization is not expected, due to the existence of a extensive
number of conserved quantities. In this regard, it has been suggested that 
the steady state of an integrable system of hard-core bosons may be described
by a generalized Gibbs ensemble (GGE) that maximizes the entropy with all 
possible constraints imposed by the existence of (infinite) integrals of 
motion~\cite{rigol2007}.

At the roots of many-body theory in condensed matter theory, the Luttinger 
model was introduced to describe a system of interacting fermions in 1D;
the approximation of considering fully linear dispersions made it possible an 
exact solution in terms of {\it bosonic} variables, e.g., fermionic 
densities~\cite{luttinger1963,lieb1965} (hence the name {\it bosonization}). 
Later, the asymptotic forms of one- and two-particle correlations in 
equilibrium were obtained by Luther and Peschel~\cite{luther1974} and 
Haldane~\cite{haldane1980,haldane1981} proposed that this model may generically
describe the low-energy properties of a broad class of systems in 1D, now 
known as Tomonaga-Luttinger (TL) liquids~\cite{giamarchi2004}.

Recently, the bosonization approach for the Luttinger model has been used to 
compute the time evolution after a quantum quench~\cite{cazalilla2006}.
According to these results, real space correlation functions evolve towards 
a steady state. Remarkably, the spatial decay of correlations is always 
governed by critical exponents, which however are different from the ground 
state ones. On the other hand, {\it generic} quantum models in 1D are expected
to give rise to thermalization, where all singularities are washed out. 
It is therefore important to test the bosonization results against numerical 
simulations of lattice models in order to better understand the non-equilibrium
dynamics of quantum one-dimensional models. Recently, a numerical investigation
of a one-dimensional lattice model of spinless fermions has been performed by 
using density-matrix renormalization group technique~\cite{karrasch2012}.
The analysis of the short-time evolution of few observables suggests that the 
bosonization predictions are indeed verified also in lattice models.
By contrast, a renormalization-group approach of the bosonized Hamiltonian in 
presence of a periodic potential has suggested that temperature effects and 
dissipation among bosonic modes may be generated, removing all
singularities~\cite{mitra2011}. In this paper, we aim at performing a detailed
comparison between the TL approach and exact calculations (by Lanczos 
diagonalizations) on a microscopic model of interacting fermions on a 1D 
lattice. Although the latter approach is limited to relatively small system 
sizes, we obtain a clear and unambiguous evidence that a TL analysis misses 
important aspects of the long-time behavior.

The paper is organized as follow: in sections~\ref{sec:bosonization} 
and~\ref{sec:correlation}, we review the basic steps of the bosonization 
technique and the results for the time-dependent correlation functions;
in section~\ref{sec:model}, we present the lattice models and the numerical
method based upon Lanczos diagonalization; in section~\ref{sec:preliminary},
we make some preliminary considerations on time evolution on finite sizes and
time averages; in section~\ref{sec:results}, we show the results; 
in section~\ref{sec:conclusions}, we draw our conclusions. Finally, in the 
appendix~\ref{app:jastrow}, we show that the exact time evolution of the TL 
model can be described by a density-density Jastrow wave function.

\section{Bosonization}\label{sec:bosonization}

The TL model is obtained from a generic Hamiltonian of interacting spinless 
fermions in a 1D lattice:
\begin{eqnarray}\label{eq:hamiltonian}
{\cal H} &=& {\cal H}_0 + {\cal H}_{\rm int}, \\
{\cal H}_0 &=& \sum_k \xi_k c^\dag_k c_k, \\
{\cal H}_{\rm int} &=& \frac{1}{2L} \sum_{k,k^\prime,q \ne 0} V(q) 
c^\dag_k c^\dag_{k^\prime} c_{k^\prime-q} c_{k+q},
\end{eqnarray}
where $\xi_k$ represents the fermionic dispersion and $L$ is the size of the
lattice. The specific form of the interaction $V(q)$ depends upon the details
of the model. In general, it is assumed that well defined values for the
forward scattering (small $q$'s) and backscattering ($q \approx \pm 2k_F$) are
given. The basic idea is to linearize the dispersion relation $\xi_k$ near the
Fermi energy, i.e., $\xi_k \approx v_F (|k|-k_F)$. It is convenient to assume
that this linear dispersion extends for all $k \in [-\infty,\infty]$. In this 
case, the particle belonging to either branch, denoted by left or right, are 
distinguishable and will be considered as two different species of fermions
(Luttinger model).~\cite{luttinger1963}

The Hamiltonian~(\ref{eq:hamiltonian}) can be written in terms of two density
operators (for left and right particles):
\begin{equation}
\rho_L(q) = \sum_{k<0} c_k^\dag c_{k+q} \;\;\;\;\; 
\rho_R(q) = \sum_{k>0} c_k^\dag c_{k+q}.
\end{equation}
Indeed, up to a constant, we obtain:
\begin{eqnarray}
{\cal H}_0 &=& \frac{2\pi v_F}{L} \sum_{q>0} 
\left[ \rho_R(q) \rho_R(-q) + \rho_L(q) \rho_L(-q) \right], \\
{\cal H}_{\rm int} &=& \frac{V}{2L} \sum_{q>0} 
\left[ \rho_R(q) + \rho_L(q) \right] \left[ \rho_R(-q) + \rho_L(-q) \right],
\end{eqnarray}
where $V=V(0)-V(2k_F)$. In the standard approach for bosonization, new boson
operators are considered:
\begin{eqnarray}\label{eq:a}
a_q &=& \sqrt{\frac{2\pi}{qL}} \rho_R(q) \;\;\;\;\; 
a_q^\dag= \sqrt{\frac{2\pi}{qL}} \rho_R(-q), \\
\label{eq:b}
b_q &=& \sqrt{\frac{2\pi}{qL}} \rho_L(-q) \;\;\;\;\; 
b_q^\dag= \sqrt{\frac{2\pi}{qL}} \rho_L(q),
\end{eqnarray}
such that $a$ and $b$ obey to the usual canonical bosonic commutation 
relations. In terms of these operators the Hamiltonian becomes: 
\begin{eqnarray}\label{eq:nonint}
{\cal H}_0 &=& v_F \sum_{q>0} q 
\left ( a_q^\dag a_q + b_q^\dag b_q \right ) \\
{\cal H}_{\rm int} &=& \frac{V}{2\pi} \sum_{q>0} q 
\left ( a_q^\dag a_q + b_q^\dag b_q + a_q b_q + b_q^\dag a_q^\dag \right ).
\end{eqnarray}
Therefore, the full Hamiltonian may be easily diagonalized by using a 
$q$-independent Bogoliubov transformation:
\begin{eqnarray}\label{eq:bogo1}
\alpha_q &=& u a_q + v b_q^\dag, \\ 
\label{eq:bogo2}
\beta_q &=& v a_q^\dag + u b_q,
\end{eqnarray}
with $u^2-v^2=1$; therefore, we can take $u=\cosh \phi$ and $v=\sinh \phi$,
with $\tanh (2\phi) = V/(2\pi v_F +V)$. We mention that $u$ and $v$ can be 
expressed in terms of the standard Luttinger parameter $K$, namely
$u=(1+K)/(2 \sqrt{K})$ and $v=(1-K)/(2 \sqrt{K})$, with
$K=(u-v)^2=\cosh 2\phi - \sinh 2\phi$.

After the Bogoliubov transformation, the TL Hamiltonian becomes
\begin{equation}\label{eq:interacting}
{\cal H} = \sum_{q>0} \epsilon_q \left( \alpha_q^\dag \alpha_q +
\beta_q^\dag \beta_q \right),
\end{equation}
where:
\begin{equation}
\epsilon_q= \left \{ \sqrt{v_F^2+\frac{v_F V}{\pi}} \right \} q = c_s q
\end{equation}
is the excitation energy for the $q$-mode.

Finally, let us indicate by $|0\rangle_{ab}$ and $|0\rangle_{\alpha\beta}$ the
vacuum states of $(a,b)$ and $(\alpha,\beta)$ bosons, which coincide with the 
ground states for the non-interacting [Eq.~(\ref{eq:nonint})] and the 
interacting [Eq.~(\ref{eq:interacting})] systems, respectively. Then, 
the ground state of the full Hamiltonian can be written as:
\begin{equation}\label{eq:gs}
|0\rangle_{\alpha\beta} \propto \exp{ \left [ -\frac{v}{u} \sum_{q>0} 
\alpha_q^\dag \beta_q^\dag \right ]} |0\rangle_{ab}.
\end{equation}

Let us now consider the time evolution of the non-interacting state
$|0\rangle_{ab}$ by using the interacting Hamiltonian ${\cal H}$, corresponding
to a quantum quench from $V=0$ to a finite value of the interaction strength. 
Interestingly, the time evolution has a very simple and instructive form, 
which will be used in the following Section. Indeed, from Eq.~(\ref{eq:gs}) 
we can write
\begin{equation}
|0\rangle_{ab} \propto \exp{ \left [ \frac{v}{u} \sum_{q>0} 
\alpha_q^\dag \beta_q^\dag \right ]} |0\rangle_{\alpha\beta},
\end{equation}
and therefore
\begin{eqnarray}
&& e^{-i{\cal H} t}|0\rangle_{ab} \propto 
\exp{\left [ \frac{v}{u} \sum_{q>0} e^{-2i \epsilon_q t} \alpha_q^\dag \beta_q^\dag \right ]} 
|0\rangle_{\alpha\beta} 
\nonumber \\ 
\label{eq:evol}
&&\propto \exp{\left [ \frac{v}{u} \sum_{q>0} \left (e^{-2i \epsilon_q t} -1 \right ) 
\alpha_q^\dag \beta_q^\dag \right ]} |0\rangle_{ab}
\end{eqnarray}

In appendix~\ref{app:jastrow}, we show that both the {\it exact} ground 
state~(\ref{eq:gs}) and the time evolution of Eq.~(\ref{eq:evol}) may be 
rewritten in term of density-density Jastrow wave functions,~\cite{jastrow1955}
which are commonly used to describe ground-state properties of correlated 
systems,~\cite{capello2005,capello2007} and more recently also out of 
equilibrium dynamics.~\cite{carleo2012}

\section{Correlation functions}\label{sec:correlation}

Here, we give a brief overview of the time-dependent correlation functions; 
our results agree with those of Ref.~\cite{cazalilla2006} for the Luttinger 
model. Let us suppose that at $t=0$ the system is set in the non-interacting 
ground state $|0\rangle_{ab}$, while for $t>0$ it evolves according to the 
interacting Hamiltonian ${\cal H}$:
\begin{equation} 
|\Phi(t) \rangle=e^{-i{\cal H}t}|0\rangle_{ab}.
\end{equation} 
Then, the bosonization technique allows one to directly compute the 
density-density correlation function:
\begin{equation}
N_q(t) = \frac{1}{L} \langle \Phi(t)| n_q n_{-q}|\Phi(t) \rangle,
\end{equation}
where $n_{\pm q} = \rho(\pm q) = \rho_L(\pm q) + \rho_R(\pm q)$. Indeed, by 
using Eqs.~(\ref{eq:a}),~(\ref{eq:b}),~(\ref{eq:bogo1}),~(\ref{eq:bogo2}) 
and~(\ref{eq:evol}), after some straightforward algebra, we obtain the simple 
expression: 
\begin{equation}\label{eq:nqtime}
N_q(t) = \frac{q}{2\pi} \left [ A + B \cos(2\epsilon_q \; t) \right ].
\end{equation}
where
\begin{eqnarray}
A &=& (u^2+v^2)(u-v)^2 = \frac{1+K^2}{2}, \\
B &=& 1-A.
\end{eqnarray}
This result shows that $N_q(t)$ does not relax to a finite limit but instead 
oscillates with a single frequency that equals twice the value of the 
excitation energy $\epsilon_q=c_s q$. By performing a Fourier transform 
(that takes into account an ultraviolet cutoff), it is possible to obtain the
$q \sim 0$ (spatially monotonic) contribution of the density-density 
correlations:
\begin{equation}
N_r(t) = n^2 - \frac{1}{4\pi^2} \Bigg[\frac{2A}{r^2} +
\frac{B}{(r+2c_s t)^2} + \frac{B}{(r-2c_s t)^2} \Bigg],
\end{equation}
which shows that, for each fixed distance $r$, $\lim_{t \to \infty} N_r(t)$ 
converges to a finite value. 
Notice that the asymptotic behavior of $N_r(t)$ coincides with the Fourier 
transform of {\it the time average} of $N_q(t)$, which equals $Aq/{2\pi}$
and differs from the known ground state behavior $Kq/{2\pi}$.

For completeness, we also report the time evolution of the $q \sim 2k_F$ 
singular contribution to the density correlations: 
\begin{equation}
N_r(t) \propto \left (\frac{4c_s^2t^2} {|r^2-4c_s^2t^2|} \right )^{2uv(u-v)^2}
\frac{\cos(2k_Fr)} {r^{2(u^2+v^2)(u-v)^2}},
\end{equation}
and the $q \sim k_F$ singularity of the one-body density matrix:
\begin{equation}
\langle \Phi(t)|c_r^\dag c_0  |\Phi(t) \rangle \propto
\left ( \frac{|r^2-4c_s^2t^2|} {4c_s^2t^2} \right )^{2u^2v^2}
\frac{\cos(k_Fr)} {r^{1+4u^2v^2}}.
\end{equation}
At fixed $r$, for $t\to\infty$ the first factor in these expressions tends to 
unity, showing that the correlation functions in real space tend again to a 
finite limit for large times. 

\section{Models and numerical methods}\label{sec:model}

Let us now define the microscopic model that will be used to make comparisons
with the bosonization predictions: a system of spinless fermions interacting 
via a repulsive short-range potential:
\begin{equation}\label{eq:ham-spinless}
{\cal H} = -J \sum_{\langle i,j \rangle} c^\dag_i c_j + H.c. 
+V \sum_{\langle i,j \rangle} n_i n_j
+V^\prime \sum_{\langle \langle i,j \rangle \rangle} n_i n_j
\end{equation}
where $\langle i,j \rangle$ and $\langle \langle i,j \rangle \rangle$ indicate 
nearest-neighbor and next-nearest-neighbor sites, respectively. $c^\dag_i$ 
($c_i$) creates (destroys) a fermion on the site $i$, and $n_i=c^\dag_i c_i$ is
the fermion density. The number of sites and fermions are denoted by $L$ and 
$N$, respectively, so that the fermion density is $n=N/L$. Periodic or 
antiperiodic boundary conditions are considered for odd or even $N$, 
respectively. In the following, we take $J=1$, as energy scale. Most of the 
calculations are done in presence of nearest-neighbor interaction $V$ only 
(which corresponds to an {\it integrable} system), similar results are also 
obtained including a finite next-nearest-neighbor $V^\prime$ (that breaks the
integrability). Notice that the system we investigate is equivalent to a model
of hard-core bosons with Hamiltonian~(\ref{eq:ham-spinless}) and periodic 
boundary conditions.

The quantum quench consists in taking an initial wave function 
$|\Psi(0)\rangle$, which is the ground state of Eq.~(\ref{eq:ham-spinless}) 
with $V=V_i$ (and $V^\prime=V^\prime_i$), and letting it evolve under the 
same Hamiltonian with $V=V_f$ (and $V^\prime=V^\prime_f$). By using the 
Lanczos method, it is possible to perform the {\it exact} time evolution of 
any initial state. Indeed, the full time interval can be split in small steps 
$\Delta t$ and the time evolution can be evaluated recursively:
\begin{equation}
|\Psi(t+\Delta t)\rangle = e^{-i{\cal H} \Delta t}|\Psi(t)\rangle.
\end{equation}
Each small-time evolution can be computed by a truncated Taylor expansion:
\begin{equation}
|\Psi(t+\Delta t)\rangle \simeq \sum_{k=0}^{k_c} \frac{(-i \Delta t)^k}{k!} 
{\cal H}^k |\Psi(t)\rangle,
\end{equation}
where the cut-off $k_c$ is chosen as to preserve energy conservation to the 
desired numerical accuracy. 

\section{Preliminary considerations}\label{sec:preliminary}

\begin{figure}
\resizebox{0.9\columnwidth}{!}{\includegraphics{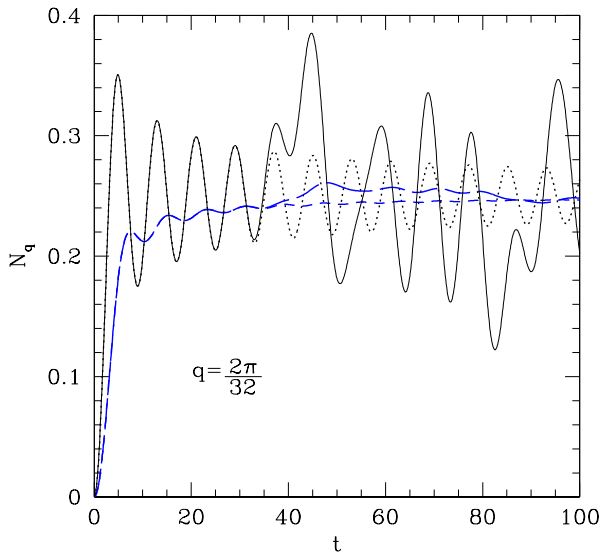}}
\caption{\label{fig:bessel}
(Color online) Time evolution of the density-density correlation $N_q(t)$
of the inhomogeneous state $|\dots,1,0,1,0,1,0,\dots\rangle$ that is evolved
with the non-interacting Hamiltonian of Eq.~(\ref{eq:ham-spinless}). Results
with $L=32$ (solid black curve) are compared with the thermodynamic limit
(dotted black curve). The time averages are also reported (blue dashed lines).}
\end{figure}

From what we have described in the previous section, by using the Lanczos 
technique, it is in principle possible to compute the time-evolution of any 
observable:
\begin{equation}
{\cal O}(t) = \langle \Psi(t)|{\cal O}|\Psi(t)\rangle.
\end{equation}
However, we must emphasize that, on finite systems, recurrence effects are 
always present and, at long times, the dynamics suffers from size effects. 
Indeed, the initial state inevitably contains excitations that, under the
time evolution, may ``travel'' all along the chain. In models where the full 
spectrum is described by non-interacting quasi-particles, interference effects
appear after the time $T \simeq L/v$, where $v$ is the velocity of the 
elementary excitations. However, for generic models, this recurrence time may 
be much larger and relies on the energy differences of the (finite-size) 
many-body spectrum. Nevertheless, although for large times ${\cal O}(t)$ may 
suffer from size effects, its average over long times is fully meaningful. 
In order to give some support on this claim, we consider a case where exact 
results can be obtained in the thermodynamic limit. In particular, we consider
the case where the density of spinless fermions is half filled (i.e., $n=1/2$) 
and the initial state has a fermion every two sites, i.e., 
$|\Psi(0) \rangle \equiv |\dots,1,0,1,0,1,0,\dots \rangle$. Then, the time 
evolution is done by the Hamiltonian~(\ref{eq:ham-spinless}) with 
$V=V^\prime=0$.  In this case, density-density correlation functions
\begin{equation}
N_r(t) = \langle\Psi(t)|n_r n_0|\Psi(t)\rangle
- \langle\Psi(t)|n_r|\Psi(t)\rangle \langle\Psi(t)|n_0|\Psi(t)\rangle
\end{equation}
can be computed analytically directly in the thermodynamic limit. This 
correlation function is translationally invariant and its Fourier transform 
is given by: 
\begin{equation}\label{eq:bessel}
N_q(t) = \frac{1}{4} \left [ 1 - J_0(8t \sin(q/2)) \right ],
\end{equation}
where $J_0(x)$ is the Bessel function of order zero. In Fig.~\ref{fig:bessel},
we compare Eq.~(\ref{eq:bessel}) with the case of $L=32$ for $q=\pi/16$. 
Although the results for $L=32$ clearly deviate from the analytical ones for 
$t \gtrsim 35$, their time average is fully meaningful and gives an excellent 
approximation of the exact outcome. Therefore, in the following we will 
consider time averages over relatively large times (up to $t=100$) to obtain 
an estimation of the long-time behavior.

\section{Numerical results}\label{sec:results}

We begin by considering the case of $V^\prime=0$. In this case, the low-energy
properties of Eq.~(\ref{eq:ham-spinless}) are well described by a (gapless) 
TL liquid, except for $n=1/2$ and $V>2$, where a (gapped) charge-density-wave
insulator is obtained.~\cite{giamarchi2004} Since we want to compare the 
numerical calculations with the TL theory, in the following, we take $n=1/4$ 
(similar results have been also obtained with other densities, e.g., $n=1/3$).
We will show the case of $V_i=0$ (and $V^\prime_i=0$), so that the initial
wave function corresponds to the case of free spinless fermions. However,
we checked that qualitatively similar results are obtained also for finite 
$V_i>0$. 

Some initial information about the time evolution can be obtained from the 
analysis of the overlaps between the initial wave function $|\Psi(0)\rangle$ 
and the eigenstates of the evolving Hamiltonian $|\Phi_n\rangle$ (with 
${\cal H} |\Phi_n\rangle = E_n |\Phi_n\rangle$) i.e., 
$c_n=\langle \Phi_n|\Psi(0)\rangle$. 
Indeed, these coefficients play an important role in the time evolution of
a generic observable ${\cal O}$:
\begin{equation}
\langle \Psi (t)|{\cal O}|\Psi (t)\rangle =
\sum_{n,n^\prime} e^{-i(E_n-E_{n^\prime})t}
c_{n^\prime}^* c_n \langle \Phi_{n^\prime}|{\cal O}|\Phi_n \rangle.
\end{equation}
This expression shows that the temporal evolution of ${\cal O}$ contains all 
the frequencies $\omega = E_n-E_{n^\prime}$ corresponding to the gaps in the 
excitation spectrum of the interacting system. Bosonization predicts a linear 
excitation spectrum at low energies that, in the special case of the observable
$N_q$ written in terms of density operators,  gives rise to a single frequency
oscillatory behavior with $\omega=2\epsilon_q$, see Eq.~(\ref{eq:nqtime}). 
The excitation spectrum of a lattice model is considerably more complex and 
curvature effects are expected to introduce further frequencies in the power 
spectrum of the time evolution of ${\cal O}$. Pure oscillatory behavior 
would correspond to a $c_n$ distribution peaked in a small energy interval 
$E_n-E_{n^\prime}$ and to matrix elements 
$\langle \Phi_{n^\prime}|{\cal O}|\Phi_n \rangle$ able to connect a single 
pair of excitations. 

\begin{figure}
\resizebox{0.9\columnwidth}{!}{\includegraphics{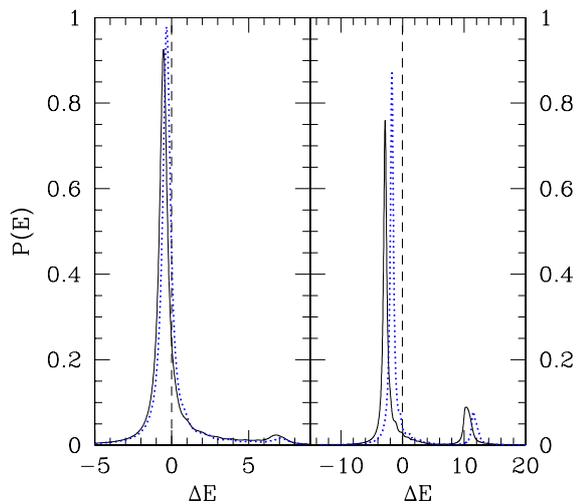}}
\caption{\label{fig:peV3V10}
(Color online) Distribution $P(E)$ of the overlaps (squared) between the
initial wave function and the eigenstates of the evolving Hamiltonian as a
function of the energy of the eigenstates $E$. Energies have been shifted,
namely $\Delta E=E-E_i$ (where $E_i$ is the energy stored into the initial
state), so that $\Delta E=0$ corresponds to the average value of the 
distribution. Lattice sizes are $L=20$ (dashed line) and $L=32$ (solid line).
A broadening of the finite size $\delta$-like peaks has been introduced.
The value of the interaction strength is $V=3$ (left panel) and $V=10$ (right
panel).}
\end{figure}

In Fig.~\ref{fig:peV3V10}, we show the results for $P(E)$ in two different 
cases with $V_f=3$ and $10$, which will be considered in the following 
discussion. The $c_n$ distribution is indeed confined in a limited energy 
interval although a significant additional peak at high energy appears for 
strong interactions. 

Let us now move to the main part of the paper and consider the density-density
correlations:
\begin{equation}
N_q(t) = \frac{1}{L} \sum_{i,j} e^{iq(R_i-R_j)} 
\langle \Psi(t)|n_i n_j|\Psi(t)\rangle.
\end{equation}
To compare the exact diagonalizations with bosonization results of
Eq.~(\ref{eq:nqtime}), we have to compute the parameters $A$ and $B$, together
with the renormalized fermionic dispersion $\epsilon_q=c_s q$. This can be 
easily done by calculating the Luttinger parameter $K$ and the velocity $c_s$,
which can be obtained either by Bethe Ansatz (for $V^\prime=0$) or numerically 
(for $V^\prime \ne 0$).~\cite{giamarchi2004}

\begin{figure}
\resizebox{0.9\columnwidth}{!}{\includegraphics{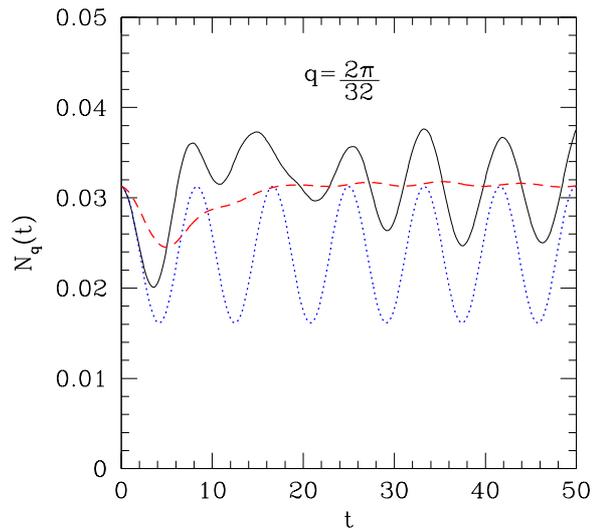}}
\caption{\label{fig:NqV3q1}
(Color online) Time evolution of the density-density correlation $N_q(t)$
at quarter filling for a quench from $V_i=0$ to $V_f=3$ and $q=2\pi/32$
(the system size is $L=32$). The time average (dashed line) and the
bosonization results (dotted line) are also shown.}
\end{figure}

\begin{figure}
\resizebox{0.9\columnwidth}{!}{\includegraphics{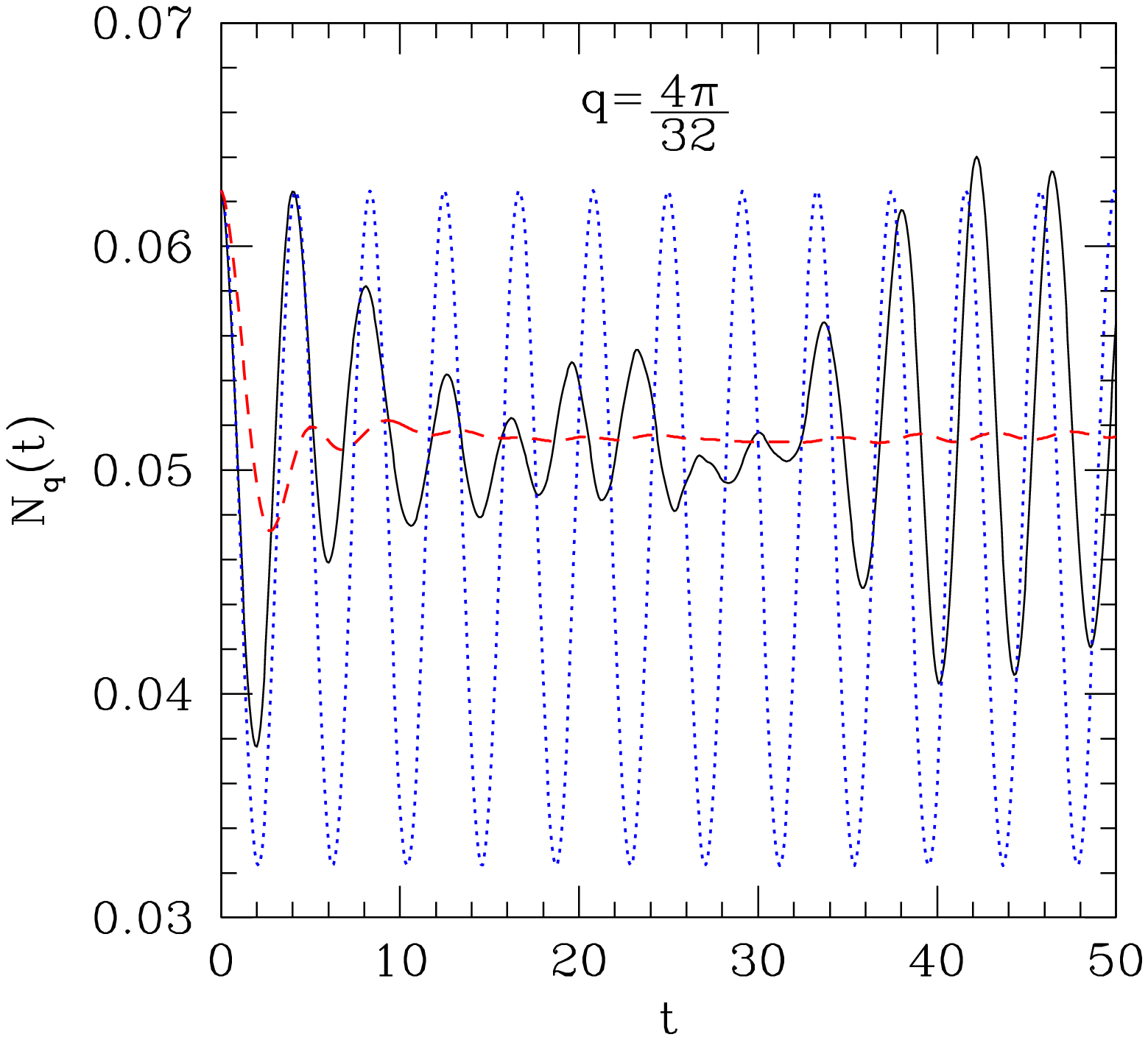}}
\caption{\label{fig:NqV3q2}
(Color online) Same as in Fig.~\ref{fig:NqV3q1} for $q=4\pi/32$.}
\end{figure}

\begin{figure}
\resizebox{0.9\columnwidth}{!}{\includegraphics{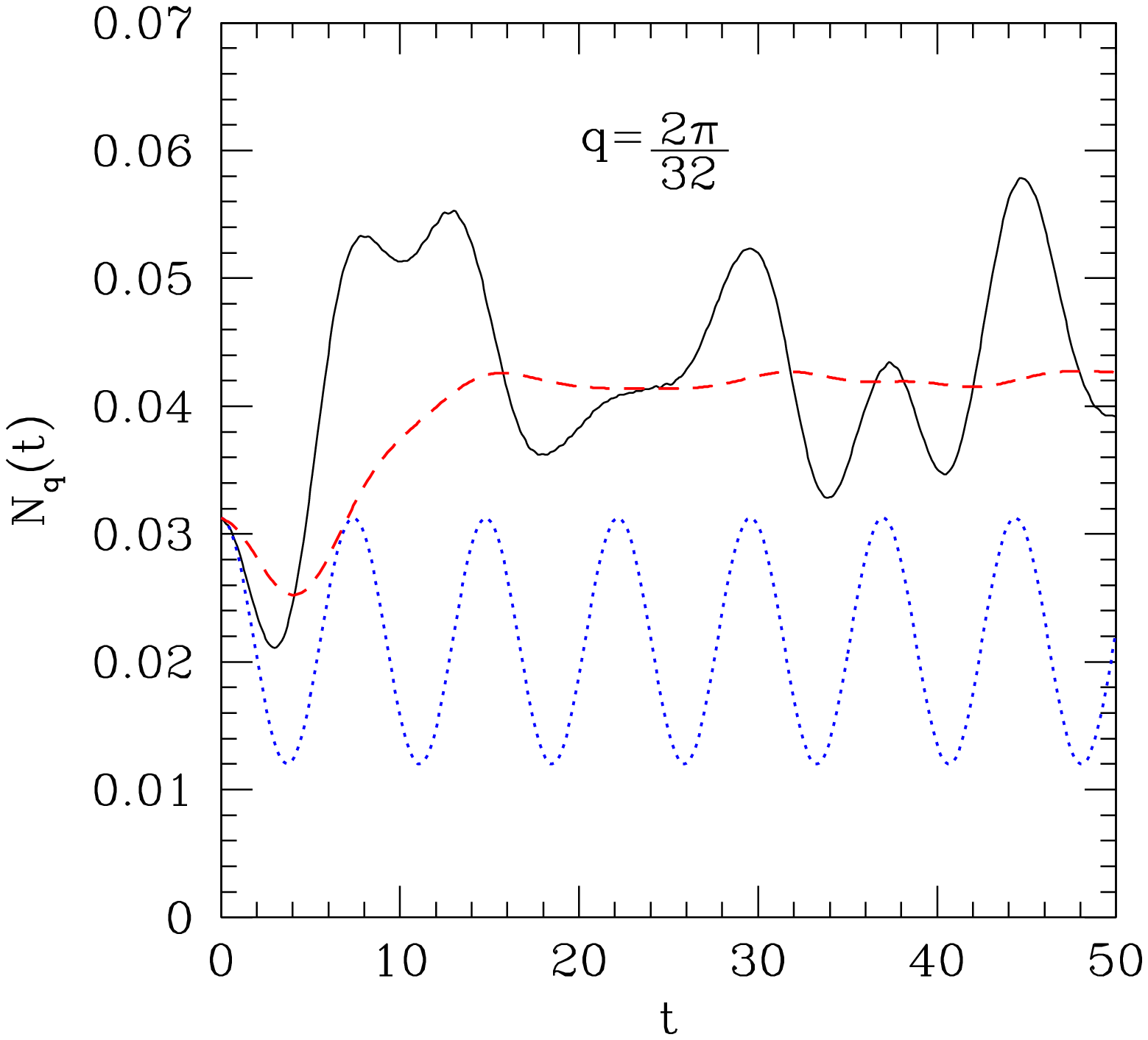}}
\caption{\label{fig:NqV10q1}
(Color online) Same as in Fig.~\ref{fig:NqV3q1} for $V_f=10$.}
\end{figure}

\begin{figure}
\resizebox{0.9\columnwidth}{!}{\includegraphics{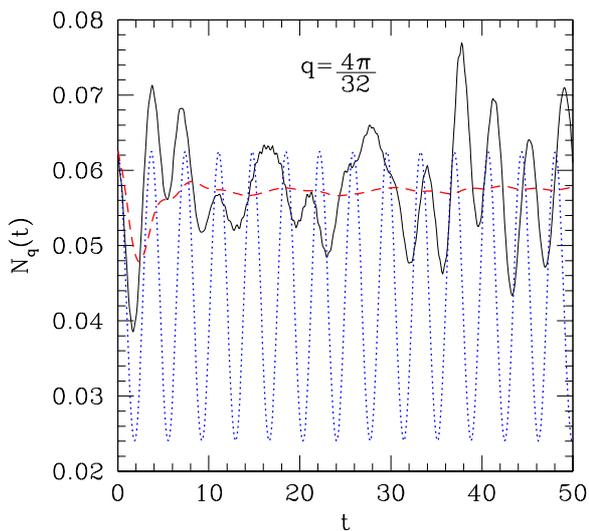}}
\caption{\label{fig:NqV10q2}
(Color online) Same as in Fig.\ref{fig:NqV10q1} for $q=4\pi/32$.}
\end{figure}

In Figs.~\ref{fig:NqV3q1},~\ref{fig:NqV3q2},~\ref{fig:NqV10q1}, 
and~\ref{fig:NqV10q2}, we report the results for $V_f=3$ and $10$ and the two
smallest non-zero momenta on the 32-site lattice, namely $q=2\pi/32$ and
$4\pi/32$, together with the bosonization prediction of Eq.~(\ref{eq:nqtime}).
The time average of the oscillating signal stabilizes at a well defined value 
for sufficiently large times, a feature also shared with the bosonization 
approach. Nevertheless, long-time averages in the TL model are always different 
from those obtained in the lattice model, where the signal shows significant 
deviations from periodicity, as a consequence of the coupling among excitation
modes. However, for small quenches (e.g., $V_f=3$) and small momenta 
(e.g., $q=2\pi/32$), after an initial transient, a stable oscillatory behavior
dominated by a single frequency is observed. In this case, the observed 
frequency agrees very well with the TL predictions. For larger values of the 
momenta and for larger quenches, the signal acquires a more complex 
periodicity.

We remark that both the amplitude and the average value of the numerical 
results considerably differ from the bosonization predictions. The discrepancy
increases for larger quenches, where, even for the smallest momenta available
in the Lanczos diagonalizations, the dynamical signal contains more than one 
frequency and the TL predictions become less and less accurate. For example, 
for $V_f=10$ and $q=2\pi/32$, the lattice model shows at least two relevant 
frequencies (see Fig.~\ref{fig:NqV10q1}), the largest one being very close to 
the TL result. Also the discrepancy between the average values grows when 
increasing the final value of the interaction strength. 

\begin{figure}
\resizebox{0.9\columnwidth}{!}{\includegraphics{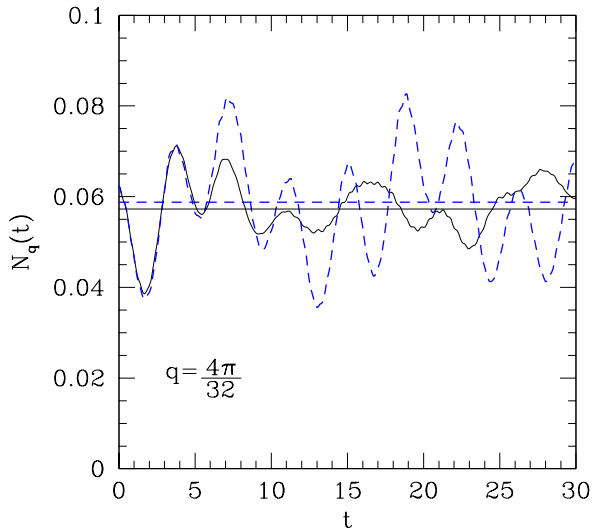}}
\caption{\label{fig:comparison}
(Color online) Time evolution of the
density-density correlation $N_q(t)$ at quarter filling for a quench from
$V_i=0$ to $V_f=3$ and $q=2\pi/32$ for $L=16$ (dashed lines) and $L=32$
(solid lines). The two averages, evaluated after a long time interval
(i.e., $t=100$), are also reported (horizontal lines).}
\end{figure}

Although the signal may have a very strong dependence on the system size for 
large times, the average value is quite stable, showing that the long-time
properties may be safely extracted from our finite-size calculations,
see Fig.~\ref{fig:comparison}. Moreover, there is some evidence that 
fluctuations around the average value decrease by increasing the cluster size,
suggesting the possibility that in the thermodynamic limit (and long times)
the signal experiences a complete damping towards its average value.

\begin{figure}
\resizebox{0.9\columnwidth}{!}{\includegraphics{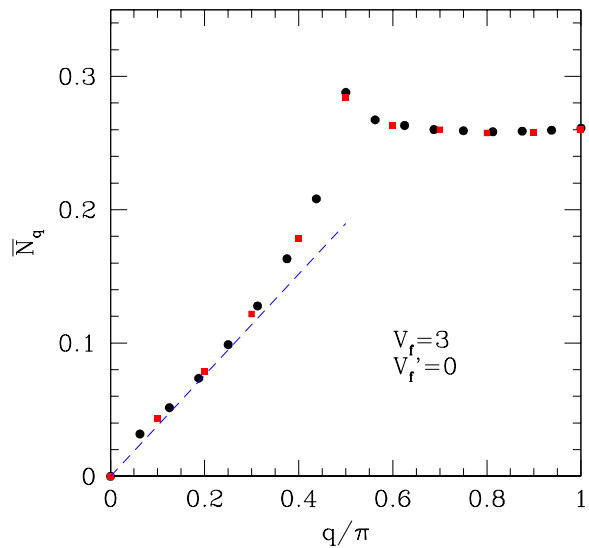}}
\caption{\label{fig:nqaveV3}
(Color online) Time average of density-density correlation $N_q(t)$ for a
quench from $V_i=0$ to $V_f=3$ for $L=20$ (squares) and $L=32$ (circles).
The dashed line shows the behavior expected from bosonization at small
momenta.}
\end{figure}

\begin{figure}
\resizebox{0.9\columnwidth}{!}{\includegraphics{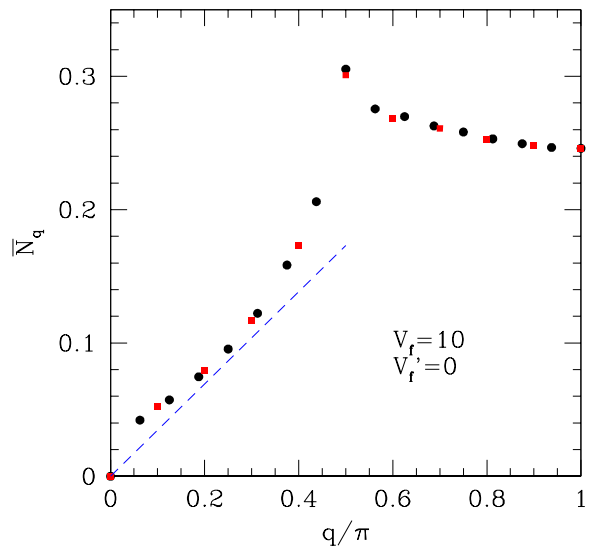}}
\caption{\label{fig:nqaveV10}
(Color online) Same as in Fig.~\ref{fig:nqaveV3} for $V_f=10$.}
\end{figure}

From these results, it is clear that bosonization misses relevant aspects 
of the dynamics after a quantum quench. More surprisingly, the long-time
average of the density-density correlations are {\it qualitatively} different
from the bosonization predictions for $q \to 0$. Indeed, in the TL theory, 
a linear behavior ${\overline N}_q=A/(2\pi)q$ (where the overbar indicates
the long-time average) is found from Eq.~(\ref{eq:nqtime}), while, the 
numerical results clearly indicate a finite limit 
${\overline N}_q= {\rm const}$ as $q \to 0$, see Figs.~\ref{fig:nqaveV3}
and~\ref{fig:nqaveV10}. The slope of ${\overline N}_q$ predicted by 
bosonization is also shown in the figures and remarkably describes the behavior 
of correlations in an intermediate range of wave-vectors. It appears that 
although the TL model does not capture an important qualitative feature of 
the long wave-length behavior of correlations, it is still able to correctly 
reproduce the physics of the asymptotic state when the size of the system is 
not exceedingly large. We will see that similar features also appear in other 
observables, like the momentum distribution. 

Although Lanczos diagonalizations are performed on finite lattices, these 
conclusions are robust against finite-size effects, as proved by the nice 
collapse of the numerical data on $L=20$ and $32$ clusters, see 
Figs.~\ref{fig:nqaveV3} and~\ref{fig:nqaveV10}. On the one hand, a 
non-vanishing long-wavelength limit of ${\overline N}_q$ is reminiscent of a 
finite temperature behavior, where singularities are washed out by thermal 
fluctuations.~\cite{notethermalization} On the other hand, our finite-size 
calculations provide some evidence for the existence of a cusp at $q=2k_F$. 
Unfortunately, due to the intrinsic statistical error induced by the time 
average procedure, the (critical) exponent related to the $q=2k_F$ singularity 
cannot be accurately determined from Lanczos data. The possibility that the 
cusps are eventually rounded in the thermodynamic limit cannot be ruled out 
either. In any case, we find that ${\overline N}_q$ cannot be suitably fitted 
by using a {\it single} effective temperature (as expected, since the model is 
integrable). 

The failure of the bosonization approach may be due to the presence of a 
perfectly linear fermionic dispersion (up to infinity or to a given cutoff) 
in the TL model. While this approximation is known to give the correct 
low-energy behavior for {\it static} properties,~\cite{affleck1989,sorella1990}
it breaks down when considering the real time evolution for long times, where 
the initial state may contain high-energy excitations, not adequately 
represented within the TL model. Moreover, bosonization describes 
{\it uncoupled} modes, which do not interact and, therefore, does not include 
the effects due to dephasing. In this regard, band curvature and finite 
bandwidth effects are expected to play a crucial role beyond the simple TL 
results.

\begin{figure}
\resizebox{0.9\columnwidth}{!}{\includegraphics{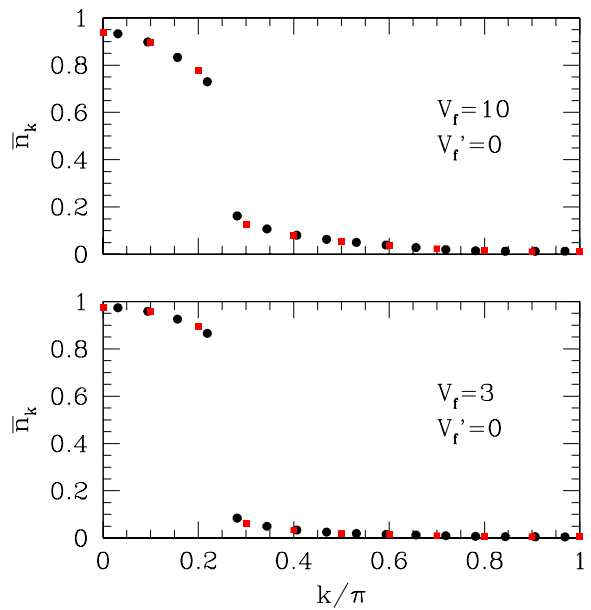}}
\caption{\label{fig:nkave}
(Color online) Time average of the momentum distribution for $L=20$ (squares)
and $L=32$ (circles) for $V_f=3$ (bottom panel) and $V_f=10$ (upper panel).}
\end{figure}

Let us now consider the momentum distribution:
\begin{equation}
n_k(t) = \frac{1}{L} \sum_{i,j} e^{ik(R_i-R_j)} 
\langle \Psi(t)|c^\dag_i c_j|\Psi(t)\rangle.
\end{equation}
The numerical results for the time average ${\overline n}_k$ are shown in 
Fig.~\ref{fig:nkave} for $V_f=3$ and $10$. On any finite-size system, the
momentum distribution shows a jump at $k_F$, which is clearly visible in our
numerical results. In the thermodynamic limit, bosonization predicts the 
occurrence of a non-analytic behavior in ${\overline n}_k$:
$\Delta {\overline n}_k \propto |k-k_F|^\alpha$, with $\alpha>0$. The exponent 
$\alpha$ could be extracted from a finite-size scaling analysis, in order
to determine the presence (for $\alpha<1$) or the absence (for $\alpha \ge 1$)
of a singularity.

\begin{figure}
\resizebox{0.9\columnwidth}{!}{\includegraphics{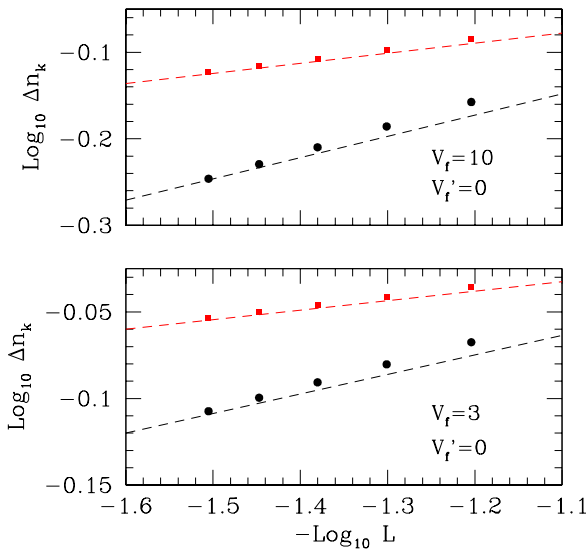}}
\caption{\label{fig:jump}
(Color online) Size scaling of the jump at $q=k_F$ in the time average of the
momentum distribution. Results for $V_f=3$ (bottom panel) and $V_f=10$
(upper panel) are reported (full circles). The results for the ground state
(squares) are also reported for comparison. The slope of the lines corresponds
to the critical exponents predicted by bosonization.}
\end{figure}

We note that the available sizes are sufficient to obtain a very accurate 
determination of the exponent for the ground state, where the exact values 
can be computed by Bethe Ansatz. On the contrary, a similar fit for the time 
averaged results appears to be more problematic, as shown in 
Fig.~\ref{fig:jump}. Indeed, a pure power-law fit of the data appears to be 
less accurate and the resulting exponent $\alpha$ increases when the fit is 
limited to the largest sizes. Nevertheless, Fig.~\ref{fig:jump} shows that the
bosonization approach provides a value for $\alpha$ roughly consistent with 
the numerical analysis, in agreement with very recent density-matrix
renormalization group calculations.~\cite{karrasch2012} However, we remark 
that a complete smoothing of the curve at larger sizes cannot be excluded by 
the Lanczos data. 

\begin{figure}
\resizebox{0.9\columnwidth}{!}{\includegraphics{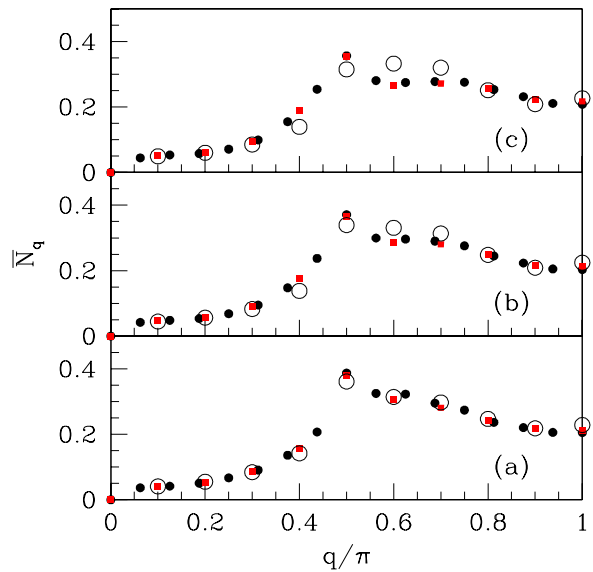}}
\caption{\label{fig:nqaveVprime}
(Color online) Time average of density-density correlation $N_q(t)$ for a
quench from $V_i=0$ to $V_f=10$ and $V^\prime_f=3$ (a), $V_f=15$ and
$V^\prime_f=4.5$ (b), and $V_f=20$ and $V^\prime_f=6$ (c). Data for $L=20$
(squares) and $L=32$ (circles) are shown. Thermal values corresponding to an
effective temperature that gives the correct internal energy are also shown
(empty circles).}
\end{figure}

\begin{figure}
\resizebox{0.9\columnwidth}{!}{\includegraphics{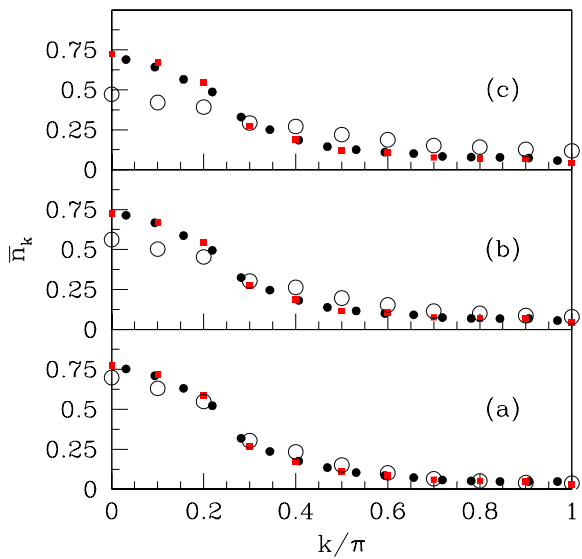}}
\caption{\label{fig:nkaveVprime}
(Color online) Same as in Fig.~\ref{fig:nqaveVprime} for the momentum
distribution $n_k(t)$.}
\end{figure}

Finally, we consider the effects of a finite $V_f^\prime$ in the final 
Hamiltonian. When sufficiently large, this term is expected to break the 
integrability conditions. In this case, thermalization should be expected on 
general grounds and we are in the position to verify whether some evidence 
for that is already visible on the lattice sizes studied by Lanczos 
diagonalization. In Figs.~\ref{fig:nqaveVprime} and~\ref{fig:nkaveVprime}, we 
report the results for ${\overline N}_q$ and ${\overline n}_k$, respectively.
Three different values of the final interaction strengths are reported. 
In all cases, ${\overline N}_q= {\rm const}$ for $q \to 0$, while a sizable 
peak is still present at $q=2k_F$. Therefore, the {\it qualitative picture} 
of the previous integrable model holds also in this case.

We also notice that, for small quenches (e.g., $V_f=10$ and $V^\prime_f=3$), 
both the density-density correlations and the momentum distribution may be 
nicely fitted by assuming a single effective temperature within the canonical 
ensemble, see Figs.~\ref{fig:nqaveVprime} and~\ref{fig:nkaveVprime}. 
Indeed, by choosing the temperature that reproduces the value of the internal
energy, we are able to obtain a quite satisfactory representation for all 
momenta, at variance with the integrable case, $V^\prime_f=0$. Of course, 
in order to show that a real thermalization takes place, one must show that 
{\it all} correlation functions are described by a {\it single} effective 
temperature. In this respect, the best way is to consider the properties of 
the (reduced) density matrix, as suggested by Poilblanc in a recent 
work.~\cite{poilblanc2011} This task is beyond the scope of the present paper,
which is centered around the comparison between the TL model and a microscopic
model on the lattice.

For larger quenches, an accurate description of these correlation functions 
in terms of a {\it single} effective temperature is less accurate
(see Figs.~\ref{fig:nqaveVprime} and~\ref{fig:nkaveVprime}) and quite 
substantial deviations from the thermal values are observed for large momenta 
in the density-density correlations and for small momenta in the momentum 
distribution. 

\section{Conclusions}\label{sec:conclusions}

In conclusion, we reported a direct comparison of the non-equilibrium
dynamics between the bosonization approach and exact diagonalizations for an
interacting model of spinless fermions. On the one hand, we showed that the 
bosonization technique does not capture few important aspects of the long-time 
behavior. In particular, a thermal-like behavior of the density-density 
correlations at small momenta is observed in the numerical calculations 
(i.e., ${\overline N}_q \simeq {\rm const}$). On the other hand, our 
numerical calculations point towards the occurrence of a singularity at the 
Fermi wavevector in the momentum distribution, as predicted by  bosonization.
Our conclusions do not crucially depend upon the presence of a 
next-nearest-neighbor interaction $V^\prime$ that breaks integrability, and
show that the critical behavior predicted by the bosonization technique
should be considered with care and may strongly depend upon the linearization
of the fermionic band. The numerical analysis for non-integrable models is 
consistent with thermalization in the thermodynamic limit, however, significant
size effects hamper the possibility to quantitatively investigate the 
thermalization dynamics. 

We thank M. Fabrizio, G. Santoro and A. Silva for stimulating discussions.
E.C. also thanks SNSF (Division II, MaNEP).

\appendix

\section{The Jastrow wave function}\label{app:jastrow}

In this Appendix, we show that the ground-state wave function~(\ref{eq:gs}) of 
the Tomonaga-Luttinger Hamiltonian may be rewritten as a Jastrow term acting
on the non-interacting ground state $|0\rangle_{ab}$, i.e., in the 
form:~\cite{tayo2008}
\begin{equation}
|0\rangle_{\alpha\beta} \propto 
\exp{ \left [ \frac{1}{L}\sum_{q>0} w_q n_{q} n_{-q} \right ]} |0\rangle_{ab},
\label{a1}
\end{equation} 
where the density operators are defined by
$n_{\pm q} = \rho(\pm q) = \rho_L(\pm q) + \rho_R(\pm q)$.
Most importantly, we also show that the {\it exact} time evolved 
state~(\ref{eq:evol}) can be also written as a (time-dependent) Jastrow term 
acting on $|0\rangle_{ab}$.

Let us start with the ground state. In order to find out the expression of the
pseudo-potential $w_q$, on the one hand, we write $|0\rangle_{\alpha\beta}$ 
by using Eqs.~(\ref{eq:bogo1}) and~(\ref{eq:bogo2}):
\begin{equation}
|0\rangle_{\alpha\beta} 
\propto \exp{ \left [ -\frac{v}{u} \sum_{q>0} 
(u a_q^\dag + v b_q)(v a_q + u b_q^\dag) \right ]} |0\rangle_{ab}.
\label{a2}
\end{equation}
On the other hand, we use the fact that:
\begin{equation}
n_{q} n_{-q} = \frac{qL}{2\pi} 
(a_q a_q^\dag + b_q^\dag b_q + a_q b_q + b_q^\dag a_q^\dag).
\label{a3}
\end{equation}
Since, in a Tomonaga-Luttinger liquid, different modes labeled by $q>0$ are
not coupled, we can focus our attention on a generic mode, dropping the label
$q$. We want to see whether, for a suitable choice of the amplitude $w_q$, 
the expression in Eq.~(\ref{a1}) coincides with that of  Eq.~(\ref{a2}). 
A solution exists if it is possible to define an amplitude $f$ such that the 
following equality holds:
\begin{equation}\label{eq:idem}
e^{f (a^\dag a + b^\dag b + a b + b^\dag a^\dag)} |0\rangle_{ab} \propto
e^{x (a a^\dag + b^\dag b +\nu a b + \frac{1}{\nu} b^\dag a^\dag)} 
|0\rangle_{ab} 
\end{equation}
for all choices of $x$ and $\nu$. In the case of interest, due to 
Eq.~(\ref{a2}) we have $x=-v^2$, $\nu=v/u$, while Eqs.~(\ref{a1}), 
and~(\ref{a3}) give $f=w_q q/(2\pi)$. Via some lengthy algebra it is possible 
to show that Eq.~(\ref{eq:idem}) is satisfied by the choice:
\begin{equation}\label{eq:idem2}
w_q = -\frac{2\pi v}{q (u-v)},
\end{equation}
in agreement with the results of Ref.~\cite{tayo2008}. 
The explicit form of the Jastrow pseudo-potential shows a long-range 
(density-density) repulsion, with a logarithmic decay in real space:
\begin{equation}
|0\rangle_{\alpha\beta} \propto \exp{ \left [ -\frac{2\pi}{L}\sum_{q>0} 
\frac{v}{q(u-v)} n_q n_{-q} \right ]} |0\rangle_{ab}.
\end{equation}

Let us now consider the time evolution and show that the time evolved wave 
function can be written as a time-dependent Jastrow term applied to the 
non-interacting state. This can be easily proved by considering 
Eq.~(\ref{eq:idem}) with $x=v^2 (e^{-2i\epsilon_q t} -1)$, $\nu=v/u$, and 
$f=w_q(t) q/(2\pi)$. The final result reads as:
\begin{equation}
e^{-i{\cal H} t}|0\rangle_{ab} \propto 
\exp{ \left [ \frac{1}{L}\sum_{q>0} w_q(t) n_q n_{-q} \right ]} |0\rangle_{ab},
\end{equation}
with
\begin{equation}
w_q(t) = \frac{2\pi v}{q(u-v)} \left [
\frac{u(e^{-2i\epsilon_q t} -1)} {u+ve^{-2i\epsilon_q t}} \right ].
\end{equation}
Therefore, we obtain the important result that the {\it exact} time 
evolution of $|0\rangle_{ab}$ under the action of ${\cal H}$ can be written
as a Jastrow wave function, with a complex time-dependent pseudo-potential 
$w_q(t)$.


\begin{thebibliography}{99}
\bibitem{notethermalization} We stress the fact that the model with 
   $V^\prime=0$ is integrable and, therefore, a complete thermalization is not
   possible. Nevertheless, some aspects of a thermal-like behavior can be 
   seen in some correlation functions.
\end{thebibliography}
\end{document}